\documentstyle[12pt]{article}
\advance\textheight by \topskip
\makeatletter

\@addtoreset{equation}{section}
\makeatother

\newcommand{\BE}{\begin{equation}}
\newcommand{\EE}{\end{equation}}
\newcommand{\BA}{\begin{eqnarray}}
\newcommand{\EA}{\end{eqnarray}}

\def\no{\nonumber}
\def\bi{\bibitem}

\def\ap{\alpha ' }
\def\B{\beta}
\def\BH{\beta_{\scriptscriptstyle H}}
\def\th{\vartheta}
\def\T2{|T|^2}
\def\TM{{\bf T}}
\def\Eb{\overline{E}}
\def\p{\partial}
\def\a{\alpha}
\def\D#1{D#1-$\overline{\textrm{D#1}}$}
\def\NS9{NS9B-$\overline{\textrm{NS9B}}$}
\def\Cm{C_{\scriptscriptstyle -1}}
\def\Dm{D_{\scriptscriptstyle -1}}
\def\Ch{C_{\scriptscriptstyle - \frac{1}{2}}}
\def\Dh{D_{\scriptscriptstyle - \frac{1}{2}}}
\def\v#1{{\cal V}_{#1}}
\def\vp{{\cal V}_{p}}
\def\A{A}
\def\db{\overline{d}}
\def\Bn{\beta_{{\mbox{\boldmath{${\scriptstyle n}$}}}}}
\def\DD{D-brane$-$anti-D-brane}

\input epsf

\begin{document}

\rightline{YITP-04-12}
\rightline{hep-th/0403078}

\vspace{.8cm}
\begin{center}
{\large\bf Finite Temperature Systems of Brane-Antibrane Pairs

and Non-BPS D-branes
}

\vskip .9 cm

{\bf Kenji Hotta,}
\footnote{E-mail address: khotta@yukawa.kyoto-u.ac.jp}

Yukawa Institute for Theoretical Physics, Kyoto University, Kyoto 606-8502, JAPAN
\vskip 1.5cm

\end{center}
\vskip .6 cm
\centerline{\bf ABSTRACT}
\vspace{-0.7cm}
\begin{quotation}

We investigate the thermodynamic properties of {\DD} pairs and non-BPS D-branes on the basis of boundary string field theory. We calculate the finite temperature effective potential of $N$ {\DD} pairs in a non-compact background and in a toroidal background. In the non-compact background case, a phase transition occurs slightly below the Hagedorn temperature, and the {\D{9}} pairs become stable. Moreover, the total energy at the critical temperature is a decreasing function of $N$ as long as the 't Hooft coupling is very small. This leads to the conclusion that a large number $N$ of {\D{9}} pairs are created simultaneously near the Hagedorn temperature. In the toroidal background case ($M_{1,9-D} \times T_{D}$), a phase transition occurs only if the {\D{p}} pair is extended in all the non-compact directions, as long as the 't Hooft coupling is very small. The total energy at the critical temperature also decreases as $N$ increases. We also calculate the finite temperature effective potential of non-BPS D-branes, and we obtain similar results. Then, we consider the thermodynamic balance between open strings on these branes and closed strings in the bulk in the ideal gas approximation, and conclude that the total energy is dominated by the open strings.

\end{quotation}

\normalsize
\newpage

\section{Introduction}
\label{sec:Intro}

Understanding the properties of unstable D-brane systems, such as coincident {\DD} pairs and non-BPS D-branes \cite{nonBPSD}, has been a subject of much interest (for a review see, e.g., Ref. \cite{Ohmori}). Type IIB string theory contains {\DD} pairs of odd dimension and non-BPS D-branes of even dimension, whereas type IIA string theory contains {\DD} pairs of even dimension and non-BPS D-branes of odd dimension. The spectrum of open strings on these unstable branes contains a tachyon field $T$. In such a brane configuration, we have $T=0$, and the potential of this tachyon field has a local maximum at $T=0$. If we assume that the tachyon potential has a non-trivial minimum, it is hypothesized that the tachyon falls into it. Sen conjectured that the tensions of these branes and the negative potential energy of the tachyon exactly cancel at the potential minimum \cite{Senconjecture}. This implies that these unstable brane systems disappear at the end of the tachyon condensation.

If tachyon condensation occurs in a topologically non-trivial way, there remain some topological defects, such as kinks and vortices. We can identify these topological defects as lower-dimensional D-branes and the topological charge as the Ramond-Ramond charge of the resulting D-branes \cite{nonBPSD} \cite{Senconjecture}. These D-brane charges can be classified using K-theory \cite{Ktheory1} \cite{Ktheory2}. In particular, we can realize all the D-branes through tachyon condensation from the spacetime-filling branes, such as {\D{9}} pairs and non-BPS D9-branes.

In recent years, significant effort has been devoted to studying the time development of the tachyon field as a ``rolling tachyon" \cite{roll}. The creation of closed strings in the rolling tachyon background has also been investigated recently \cite{closedcre}. Thus it is natural to consider the problem of how to realize the initial conditions of a rolling tachyon. In other words, we wish to consider how the tachyon field should be placed on the top of the potential. If there is a situation in which $T=0$ is first the stable potential minimum, namely, the {\DD} pair or the non-BPS D-brane is stable, we can realize such a configuration. Then, if it becomes unstable, the tachyon will roll down to the potential minimum. We have pointed out  in the previous papers that there are cases that these unstable branes become stable at very high temperatures \cite{Hotta4} \cite{Hotta5}. In particular, the spacetime-filling {\D{9}} pair becomes stable at sufficiently high temperature in all cases we have studied.

We have considered finite temperature {\DD} systems \cite{Hotta4} \cite{Hotta5}\footnote{For related discussions see Refs. \cite{LowTtach} and \cite{Huang}.} in the framework of boundary string field theory (BSFT) \cite{BSFT1} \cite{BSFT2}. This theory is based on the Batalin-Vilkoviski formalism \cite{BV}. For a superstring, the solution of the classical master equation is given by \cite{tachyon2} \cite{tachyon3} \cite{TakaTeraUe}
\BE
  S_{eff} = Z_0,
\label{eq:SoZo}
\EE
where $S_{eff}$ is the spacetime effective action and $Z_0$ is the disk partition function of the two-dimensional world sheet theory. If we calculate this solution in a constant tachyon background in the case of a single {\D{p}} pair in type II string theory, we obtain the tachyon potential \cite{tachyon2} \cite{TakaTeraUe} \cite{tachyon1} 
\BE
  V(T) = 2 \tau_p \vp \exp (-8 \T2).
\EE
Here, $T$ is a complex scalar tachyon field, $\vp$ is the $p$-dimensional volume of the system that we are considering, and $\tau_p$ is the tension of a Dp-brane, which is defined by
\BE
  \tau_p = \frac{1}{(2 \pi)^p {\ap}^{\scriptscriptstyle \frac{p+1}{2}} g_s},
\label{eq:tension}
\EE
where $g_s$ is the coupling constant of strings. The minimum of this potential is at $|T| = \infty$, and it satisfies Sen's conjecture \cite{Senconjecture}. In order to calculate the free energy by using the Matsubara method in the ideal gas approximation, we must calculate the one-loop amplitude. We have assumed that the relation (\ref{eq:SoZo}) also holds if we include the higher loop correction of the world sheet theory, namely,
\BE
  S_{eff} = Z,
\label{eq:SZ}
\EE
where $Z$ is partition function to all orders of the two-dimensional world sheet theory. If we consider the one-loop amplitude based on BSFT, we are confronted with the problem of the choice of the Weyl factors \cite{1loopAO} \cite{1loopann} \cite{1loopsym} \cite{1loop}, as we will explain in the next section. Andreev and Oft have proposed one choice \cite{1loopAO}, and we have computed the one-loop free energy using it \cite{Hotta4} \cite{Hotta5}. Then, we computed the finite temperature effective potential of {\DD} systems near the Hagedorn temperature and discovered that there are cases in which $T=0$ becomes a stable minimum near the Hagedorn temperature, although it is a local maximum at zero temperature. The purpose of this paper is to generalize the above-mentioned calculation to finite temperature systems with multiple {\DD} pairs and multiple non-BPS branes and to discuss their thermodynamic properties.

This paper is organized as follows. In \S \ref{sec:thermo} we review the thermodynamic properties of strings near the Hagedorn temperature and those of open strings on the {\DD} system \cite{Hotta4} \cite{Hotta5}. In \S \ref{sec:Npair} we evaluate the finite temperature effective potential of $N$ {\DD} pairs in a non-compact background and in a toroidal background ($M_{1,9-D} \times T_{D}$) on the basis of BSFT. In \S \ref{sec:nonBPS} we generalize this {\DD} case to the non-BPS D-brane case. Then, in \S \ref{sec:balance} we study the thermodynamic balance of open strings on these branes and closed strings in the bulk. In \S \ref{sec:annulus} we consider the choice of the Weyl factors by investigating the case of another choice as an example. Finally, \S \ref{sec:conclusion} contains conclusions and discussions.

\section{Thermodynamics of a String Gas near the Hagedorn Temperature}
\label{sec:thermo}

In this section, we review the thermodynamic properties of an ideal gas of strings near the Hagedorn temperature. We summarize the basic properties of a string gas in \S \ref{sec:stringgas}, and we summarize the results of previous papers \cite{Hotta4} \cite{Hotta5} with regard to the thermodynamics of a {\DD} system in \S \ref{sec:single}.

\subsection{Thermodynamics of a String Gas}
\label{sec:stringgas}

In the canonical ensemble method, all the statistical variables are derived from the partition function. The partition function $Z(\B)$ can be obtained from the free energy $F( \B )$ as
\BE
  Z(\B) = \exp [ - \B F( \B ) ].
\label{eq:partitionfree}
\EE
The free energy can be calculated using the Matsubara method; that is, the free energy is given by the path integral of connected graphs of strings on the space where the Euclidean time direction is compactified with a circumference equal to the inverse temperature, $\B$. Thus, it seems that we can calculate the statistical variables of strings by using the canonical ensemble method. However, we cannot trust the canonical ensemble method near the Hagedorn temperature for the following reason \cite{Efura}. The partition function $Z(\B)$ is given by the Laplace transformation of the density of states $\Omega (E)$:
\BE
  Z(\B) = \int_{0}^{\infty} dE \ \Omega (E) e^{- \B E}.
\label{eq:Lap}
\EE
We can expect that the canonical ensemble method gives the same values as the microcanonical ensemble method if the integrand has a sharp peak. However, the density of states of strings behaves as
\BE
  \Omega (E) \sim e^{\BH E}
\EE
for large energy $E$, where $\BH$ is the inverse of the Hagedorn temperature,
\BE
  \BH = 2 \pi \sqrt{2 \ap},
\label{eq:BH}
\EE
and the integrand has no sharp peak near the Hagedorn temperature. For this reason, we must compute the finite temperature effective potential by using the microcanonical ensemble method, which is more fundamental than the canonical ensemble method in the sense that it is derived directly from ergodic theory. The energy $E$ is more useful than $\B$ to analyze the string gas at high temperature, because $\B$ remains almost equal to $\BH$, varying little even if $E$ varies considerably.

\begin{figure}
\begin{center}
$${\epsfxsize=6.5 truecm \epsfbox{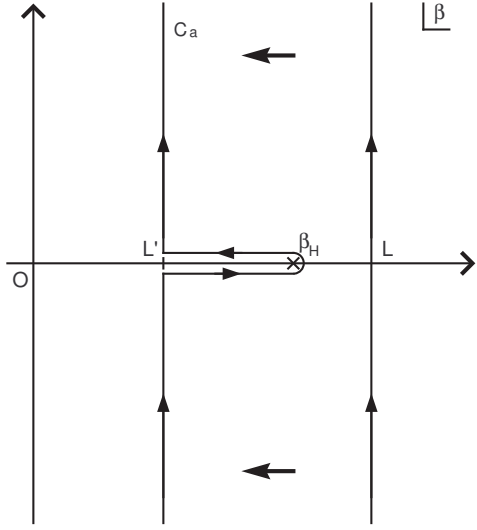}}$$
\caption{Complex $\beta$ plane.}
\label{fig:com1}
\end{center}
\end{figure}

All the statistical variables are derived from the density of states in the microcanonical ensemble method. The density of states, $\Omega (E)$, can be obtained from the inverse Laplace transformation of $Z(\B)$ as
\BE
  \Omega (E) = \int_{L-i \infty}^{L+i \infty}
    \frac{d \B}{2 \pi i} Z(\B) e^{\B E},
\label{eq:inLap}
\EE
where we must treat $\B$ as a complex variable in this case. We must choose the constant $L$ such that the path of integration in the complex $\B$-plane lies to the right side of all the singularities of the integrand. Then, we can deform the contour to the left so as to pick up the singularities, as sketched in Figure \ref{fig:com1}. The density of states has been derived from this formula in the case of closed strings \cite{Tan2} \cite{Hotta2}, in the case of open strings on D-branes \cite{Thermo}, and in the case of open strings on a single {\D{p}} pair \cite{Hotta4} \cite{Hotta5}.

In order to calculate the density of states from (\ref{eq:inLap}), we must investigate the singular part of the free energy, because the partition function is obtained from the free energy as (\ref{eq:partitionfree}). Let us separate the free energy into regular and singular parts, as
\BE
  F(\B) = F_{reg} (\B) + F_{sing} (\B),
\EE
and expand the regular part in $(\B - \BH)$ as
\BE
  - \B F_{reg} \simeq \lambda_0 \vp - \sigma_0 \vp (\B - \BH).
\label{eq:Freg}
\EE
It is easy to see that $\lambda_0$ is a constant with dimensions of inverse volume, and $\sigma_0$ is a constant with dimensions of energy density. We call $\sigma_0$ the Hagedorn energy density. Then, we obtain the density of states as
\BE
  \Omega (T,E) \simeq e^{\BH E + \lambda_0 \vp}
    \int_{C_a} \frac{d \B}{2 \pi i}
      \exp \left[ (\B - \BH) \Eb
        + F_{sing} (\B) \right],
\label{eq:dos}
\EE
where $\Eb \equiv E - \sigma_0 \vp$ and the contour $C_a$ is taken as sketched in Figure \ref{fig:com1}. A general property of the singular part of the free energy of strings is that its leading singularity is at $\B = \BH$. We call this the Hagedorn singularity. If we obtain the explicit form of $\Omega (E)$ from (\ref{eq:inLap}), the entropy $S(E)$ can be derived from $\Omega (E)$ as
\BE
  S(E) = \ln \Omega (E) \delta E,
\label{eq:Sdef}
\EE
where $\delta E$ represents the energy fluctuation, and the inverse temperature $\B$ is given by the partial derivative of $S(E)$ with respect to $E$:
\BE
  \B = \frac{\p S}{\p E}.
\label{eq:Bdef}
\EE
The finite temperature effective potential of the {\D{p}} system can be calculated from these variables as
\BE
  V(T,E) = V(T) - \B^{-1} S.
\label{eq:Vdef}
\EE
From this, we can study the stability of a system at finite temperature. As we see below, there are cases in which unstable brane configuration at zero temperature becomes stable at finite temperature.

\subsection{Finite Temperature Effective Potential for Single D-brane$-$anti-D-brane Pair}
\label{sec:single}

In this subsection we review the results of our previous papers \cite{Hotta4} \cite{Hotta5}. As explained in the previous subsection, we must first compute the free energy of open strings in order to compute the finite temperature effective potential. It can be obtained by calculating the amplitude in the space where the Euclidean time direction is compactified with a circumference equal to the inverse temperature $\B$. Let us employ the weak coupling approximation and treat strings as an ideal gas. We take into account only the one-loop amplitude, with the corresponding cylinder world sheets winding in the Euclidean time direction at least once. 

The one-loop amplitude of open strings in a {\D{p}} system in the Minkowski spacetime has been investigated on the basis of BSFT \cite{1loopAO} \cite{1loopann} \cite{1loopsym} \cite{1loop}. The one-loop world sheet of an open string has the annulus topology and two boundaries. When we compute the one-loop amplitude of open strings in a {\D{p}} system, we are confronted with an ambiguity in the choice of the Weyl factors of the two boundaries of this world sheet. This is because the conformal invariance is broken by the boundary terms in BSFT. At the tree level with a disk world sheet, we do not have such a problem, because it has only one boundary.

Andreev and Oft have proposed the following form of the one-loop amplitude \cite{1loopAO} in the Minkowski spacetime in type II string theory on the basis of the principle that its low energy part should coincide with that of the tachyon field model \cite{tachyon2} \cite{TakaTeraUe} \cite{tachyon1}. If we restrict ourselves to a constant tachyon field, which we denote by $T$, it is given by
\BA
  {\it Z_{1}} &=& \frac{16 \pi^4 i \vp}{(2 \pi \ap)^{\frac{p+1}{2}}}
    \int_{0}^{\infty} \frac{d \tau}{\tau}
      (4 \pi \tau)^{- \frac{p+1}{2}} e^{-4 \pi \T2 \tau} \no \\
  && \times \left[
    \left( \frac{\th_3 (0 | i \tau)}{{\th_1}' (0 | i \tau)} \right)^4
      - \left(\frac{\th_2 (0 | i \tau)}{{\th_1}' (0 | i \tau)} \right)^4 \right],
\label{eq:AOoneloopAmp}
\EA
where $\vp$ is the volume of the system that we are considering. This amplitude can be obtained straightforwardly by choosing the boundary action as
\BE
  S_b = \int_{0}^{2 \pi \tau} d \sigma_0 \int_{0}^{\pi} d \sigma_1
    [ \T2 \delta (\sigma_1) + \T2 \delta (\pi - \sigma_1) ].
\EE
Let us call this the `cylinder boundary action'. This action is natural in the sense that both sides of the cylinder world sheet are treated on an equal footing in this case. Here we are considering only the coincident {\D{p}} system, and we do not treat the case of a parallel brane and antibrane pair with a finite distance \cite{Huang} \cite{parallel} in this paper. 

If we compute the one-loop amplitude in the space where the Euclidean time direction is compactified with a circumference equal to the inverse temperature $\B$, we obtain the free energy. The result is \cite{Hotta4}

\newpage

\BA
  F (T, \B) &=& - \frac{16 \pi^4 \vp}{(2 \pi \ap)^{\frac{p+1}{2}}}
    \int_{0}^{\infty} \frac{d \tau}{\tau}
      (4 \pi \tau)^{- \frac{p+1}{2}} e^{-4 \pi \T2 \tau} \no \\
  && \hspace{2cm} \times \left[ \left(\frac{\th_3 (0 | i \tau)}
    {{\th_1}' (0 | i \tau)} \right)^4
      \left( \th_3 \left( 0 \left| \frac{i \B^2}{8 \pi^2 \ap \tau} \right.
        \right) -1 \right) \right. \no \\
  && \hspace{4cm} - \left.
    \left( \frac{\th_2 (0 | i \tau)}{{\th_1}' (0 | i \tau)} \right)^4
      \left( \th_4 \left( 0 \left| \frac{i \B^2}{8 \pi^2 \ap \tau} \right.
        \right) -1 \right) \right].
\label{eq:freenoncompact}
\EA
This is the free energy in a non-compact flat background. This free energy can be obtained from the proper time form of the free energy \cite{Pol}, which is given by
\BA
  F (\B) &=& - \frac{\vp}{(2 \pi \ap)^{\frac{p+1}{2}}}
    \int_{0}^{\infty} \frac{d \tau}{\tau}
      (4 \pi \tau)^{- \frac{p+1}{2}} \sum_{{M_{NS}}^2} \sum_{r=1}^{\infty}
        \exp \left( -2 \pi \ap {M_{NS}}^2 \tau
          - \frac{r^2 \B^2}{8 \pi \ap \tau} \right) \no \\
  && + \frac{\vp}{(2 \pi \ap)^{\frac{p+1}{2}}}
    \int_{0}^{\infty} \frac{d \tau}{\tau}
      (4 \pi \tau)^{- \frac{p+1}{2}} \sum_{{M_{R}}^2} \sum_{r=1}^{\infty}
        (-1)^r \exp \left( -2 \pi \ap {M_{R}}^2 \tau
          - \frac{r^2 \B^2}{8 \pi \ap \tau} \right), \no \\
\label{eq:propertime}
\EA
\begin{figure}
\begin{center}
$${\epsfxsize=6.5 truecm \epsfbox{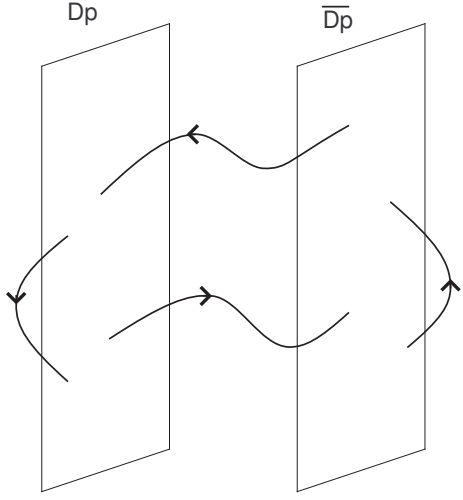}}$$
\caption{Here, for clarity, we display the brane and antibrane as if they were separated but, in fact, we consider the case of a coincident {\DD} pair.}
\label{fig:Dp}
\end{center}
\end{figure}
by multiplying by $2$ and by substituting the mass spectra
\BA
  {M_{NS}}^2
    &=& \frac{1}{\ap} \left( N_B + N_{NS} + 2 \T2 - \frac{1}{2} \right),
\label{eq:massNS} \\
  {M_{R}}^2
    &=& \frac{1}{\ap} \left( N_B + N_{R} + 2 \T2 \right),
\label{eq:massR}
\EA
where $M_{NS}$ and $M_{R}$ are the mass of the Neveu-Schwarz and Ramond sectors, respectively, and $N_B$, $N_{NS}$ and $N_{R}$ are the oscillation modes of the boson, Neveu-Schwarz fermion and Ramond fermion, respectively. This can be interpreted within open string theory as follows. As depicted in Figure \ref{fig:Dp}, there are four types of open strings, namely, two types of strings whose two ends attach to the same brane and two types of strings that are stretched between the Dp-brane and the $\overline{\textrm{Dp}}$-brane. If we denote the open strings extending from a Dp-brane to a $\overline{\textrm{Dp}}$-brane as (Dp,$\overline{\textrm{Dp}}$), the others are denoted as ($\overline{\textrm{Dp}}$,Dp), (Dp,Dp) and ($\overline{\textrm{Dp}}$,$\overline{\textrm{Dp}}$). We must impose the GSO projection for (Dp,Dp) and ($\overline{\textrm{Dp}}$,$\overline{\textrm{Dp}}$) and the opposite GSO projection for (Dp,$\overline{\textrm{Dp}}$) and ($\overline{\textrm{Dp}}$,Dp). Therefore, instead of employing the GSO projection, we may multiply by an overall factor of $2$.

The mass spectra given in (\ref{eq:massNS}) and (\ref{eq:massR}) are shifted by $2 \T2 / \ap$ from the mass spectra on the {\D{p}} system at $T=0$. This shift of mass is reminiscent of the mini-superspace approximation in S-brane thermodynamics proposed by Maloney, Strominger and Yin \cite{SbraneThermo}. Their argument is based on boundary conformal field theory (BCFT), and the boundary action is chosen as that corresponding to the case of a time dependent tachyon field.\footnote{The boundary action in the bosonic string case is chosen as
\BA
  S_b = \lambda \int d \tau \cosh \frac{X^0 (\tau)}{\sqrt{\ap}}, \no
\EA
where $\lambda$ is a constant real number \cite{SbraneThermo}.} They treat it as a time dependent mass and compute the creation rate of the open strings. Our model is related to their model in the sense that the boundary term is treated as a shift of the mass spectra.

We can also compute the free energy in a flat background with some toroidally compactified directions in a similar way. We assume that $D$-dimensional space is compactified and that the rest of the $(9-D)$-dimensional space is left uncompactified (namely, $M_{1,9-D} \times T_{D}$). We call this background the `toroidal background'. We also assume that the {\D{p}} system extends in the $d$-dimensional space in the non-compact direction and in the $(p-d)$-dimensional space in the toroidal direction. The {\D{p}} system extends in all non-compact direction if $D+d=9$. The free energy is given by
\BA
  F (T, \B ,R) &=& - \frac{16 \pi^4 \v{d}}{{\BH}^{d+1}}
    \int_{0}^{\infty} \frac{d \tau}{\tau^{\frac{d+3}{2}}}
      e^{-4 \pi \T2 \tau} \prod_{I=1}^{p-d} \prod_{i=p-d+1}^{D}
        \th_3 \left( 0 \left| \frac{2 i \ap \tau}{{R_I}^2} \right. \right)
          \th_3 \left( 0 \left| \frac{2 i {R_i}^2 \tau}{\ap} \right. \right)
            \no \\
  && \hspace{2cm} \times \left[ \left(\frac{\th_3 (0 | i \tau)}
    {{\th_1}' (0 | i \tau)} \right)^4
      \left( \th_3 \left( 0 \left| \frac{i \B^2}{{\BH}^2 \tau} \right.
        \right) -1 \right) \right. \no \\
  && \hspace{4cm} - \left.
    \left( \frac{\th_2 (0 | i \tau)}{{\th_1}' (0 | i \tau)} \right)^4
      \left( \th_4 \left( 0 \left| \frac{i \B^2}{{\BH}^2 \tau} \right.
        \right) -1 \right) \right],
\label{eq:freecompact}
\EA
where $\v{d}$ is the $d$-dimensional volume in the non-compact directions parallel to the {\D{p}} system. The effect of compactification is reflected by the infinite product of $\th_3$-functions. We can also obtain this free energy from (\ref{eq:propertime}) by multiplying by $2$ and substituting the mass spectra
\BA
  {M_{NS}}^2
    &=& \sum_{I=1}^{p-d} \left( \frac{m_I}{R_I} \right)^2
      + \sum_{i=p-d+1}^{D} \left( \frac{n_i R_i}{\ap} \right)^2 
        + \frac{1}{\ap} \left( N_B + N_{NS} + 2 \T2 - \frac{1}{2} \right),
\label{eq:massNStorus} \\
  {M_{R}}^2
    &=& \sum_{I=1}^{p-d} \left( \frac{m_I}{R_I} \right)^2
      + \sum_{i=p-d+1}^{D} \left( \frac{n_i R_i}{\ap} \right)^2 
        + \frac{1}{\ap} \left( N_B + N_{R} + 2 \T2 \right),
\label{eq:massRtorus}
\EA
where $m_I$ denotes the momentum number in the directions parallel to the {\D{p}} system, and $n_i$ denotes the winding number in the directions transverse to it. It should be noted that these free energy and mass spectra are invariant under the T-duality transformation \cite{Tdual}
\BE
  R_I \rightarrow \frac{\ap}{R_i} \ \ \ , \ \ \ m_I \rightarrow n_i,
\label{eq:Tdualmn}
\EE
for directions parallel to the {\D{p}} system and
\BE
  R_i \rightarrow \frac{\ap}{R_I} \ \ \ , \ \ \ n_i \rightarrow m_I,
\label{eq:Tdualnm}
\EE
for directions transverse to it. From these free energies, we can compute the finite temperature effective potential by using the method presented in the previous subsection.

We now summarize the results of previous papers \cite{Hotta4} \cite{Hotta5}. In the non-compact background case, the results for the {\D{9}} case differ from those in the other cases. In the {\D{9}} case, the sign of the coefficient of the $\T2$ term of the finite temperature effective potential changes from negative to positive slightly below the Hagedorn temperature as the temperature increases. This means that a phase transition occurs near the Hagedorn temperature and the {\D{9}} pair becomes stable. On the other hand, the coefficient remains negative in the {\D{p}} case with $p \leq 8$, and thus a phase transition does not occur. This leads us to the conclusion that only a {\D{9}} pair is created near the Hagedorn temperature. In the toroidal background case, a phase transition occurs near the Hagedorn temperature only in the $D+d=9$ case, namely, the case in which the {\D{p}} pair is extended in all non-compact directions. The coefficient remains negative in the {\D{p}} case with $D+d \leq 9$, and thus a phase transition does not occur.

\section{Finite Temperature Effective Potential for Multiple D-brane$-$anti-D-brane Pairs}
\label{sec:Npair}

We have considered only a single {\DD} pair to this point. Now we generalize the calculation of the previous section to the case of multiple {\DD} pairs. We begin by reviewing the tachyon field in the case of multiple {\DD} pairs in \S \ref{sec:NTfield}. We then investigate the behavior of the free energy near the Hagedorn singularity in \S \ref{sec:Hagsing}. Finally, we calculate the finite temperature effective potential in a non-compact background and in a toroidal background in \S \ref{sec:noncompact} and \S \ref{sec:torus}, respectively.

\subsection{Tachyon Field for Multiple D-brane$-$anti-D-brane Pairs}
\label{sec:NTfield}

In the single {\DD} case, the tachyonic field is a complex scalar field. If we consider open strings on $N$ {\DD} pairs, the spectrum of the open strings contains a tachyon field in the $(N, \overline{N})$ representation of the $U(N) \times U(N)$ gauge group \cite{IIBornotIIB}. In this case, the tree level tachyon potential in BSFT can be written \cite{tachyon2} \cite{TakaTeraUe}
\BE
  V(\TM) = 2 \tau_p \vp \ {\rm Tr} \exp \left( -8 {\TM}^{\dagger} \TM \right),
\label{eq:TMNpotential}
\EE
where $\TM$ is an $N \times N$ complex matrix. This potential depends on $\TM$ only in the form ${\TM}^{\dagger} \TM$. At the one-loop level, the free energy can be written
\BA
  F (\TM, \B) &=& - \frac{16 \pi^4 \vp}{{\BH}^{p+1}}
    \int_{0}^{\infty} \frac{d \tau}{{\tau}^{\frac{p+3}{2}}}
      \ {\rm Tr} \exp \left( -2 \pi {\TM}^{\dagger} \TM \tau \right)
        \ {\rm Tr} \exp \left( -2 \pi {\TM}^{\dagger} \TM \tau \right) \no \\
  && \hspace{2cm} \times \left[ \left(\frac{\th_3 (0 | i \tau)}
    {{\th_1}' (0 | i \tau)} \right)^4
      \left( \th_3 \left( 0 \left| \frac{i \B^2}{{\BH}^2 \tau} \right.
        \right) -1 \right) \right. \no \\
  && \hspace{4cm} - \left.
    \left( \frac{\th_2 (0 | i \tau)}{{\th_1}' (0 | i \tau)} \right)^4
      \left( \th_4 \left( 0 \left| \frac{i \B^2}{{\BH}^2 \tau} \right.
        \right) -1 \right) \right].
\label{eq:Freemat}
\EA
This free energy also depends on $\TM$ only in the form ${\TM}^{\dagger} \TM$. Because ${\TM}^{\dagger} \TM$ is a hermitian matrix, it can be diagonalized by a unitary transformation. This corresponds to choosing the appropriate linear combination of branes and antibranes. Let us choose ${\TM}^{\dagger} \TM$ as
\BE
  {\TM}^{\dagger} \TM =
     \left( \begin{array}{ccccc}
      \T2  &       &       &       &   0   \\
           & \cdot &       &       &       \\
           &       & \cdot &       &       \\
           &       &       & \cdot &       \\
       0   &       &       &       &  \T2  \\
     \end{array} \right),
\EE
which corresponds to the choice that all the tachyon fields on each {\D{p}} pair condenses by the same amount. The finite temperature effective potential does not depend on the choice of $\TM$, as long as it gives above form of ${\TM}^{\dagger} \TM$. Thus we can choose $\TM$ as
\BE
  \TM =
     \left( \begin{array}{ccccc}
       T   &       &       &       &   0   \\
           & \cdot &       &       &       \\
           &       & \cdot &       &       \\
           &       &       & \cdot &       \\
       0   &       &       &       &   T   \\
     \end{array} \right).
\label{eq:TM}
\EE
Substituting (\ref{eq:TM}) into (\ref{eq:TMNpotential}) and (\ref{eq:Freemat}), we obtain
\BE
  V(T) = 2 N \tau_p \vp \exp (-8 \T2)
\label{eq:TNpotential}
\EE
and 

\newpage

\BA
  F (T, \B) &=& - \frac{16 \pi^4 N^2 \vp}{{\BH}^{p+1}}
    \int_{0}^{\infty} \frac{d \tau}{{\tau}^{\frac{p+3}{2}}}                          \exp \left( -4 \pi \T2 \tau \right) \no \\
  && \hspace{2cm} \times \left[ \left(\frac{\th_3 (0 | i \tau)}
    {{\th_1}' (0 | i \tau)} \right)^4
      \left( \th_3 \left( 0 \left| \frac{i \B^2}{{\BH}^2 \tau} \right.
        \right) -1 \right) \right. \no \\
  && \hspace{4cm} - \left.
    \left( \frac{\th_2 (0 | i \tau)}{{\th_1}' (0 | i \tau)} \right)^4
      \left( \th_4 \left( 0 \left| \frac{i \B^2}{{\BH}^2 \tau} \right.
        \right) -1 \right) \right],
\label{eq:FreeN}
\EA
respectively. The tachyon potential (\ref{eq:TNpotential}) is $N$ times that of a single {\D{p}} pair, and the free energy (\ref{eq:FreeN}) is $N^2$ times that of a single {\D{p}} pair. This free energy can be obtained from the proper time form of the free energy (\ref{eq:propertime}) by multiplying by $2N^2$ and substituting the mass spectra (\ref{eq:massNS}) and (\ref{eq:massR}). The factor of $2N^2$ comes from the fact that there are $2N^2$ types of strings with (Dp,Dp) and ($\overline{\textrm{Dp}}$,$\overline{\textrm{Dp}}$) and $2N^2$ types of strings with (Dp,$\overline{\textrm{Dp}}$) and ($\overline{\textrm{Dp}}$,Dp). The 't Hooft coupling $g_s N$ must be very small for the ideal gas approximation in this case.\footnote{If the 't Hooft coupling is very large, we must consider the black brane solution in supergravity, which corresponds to the brane-antibrane system \cite{blackbrane}. However, because we are assuming that the 't Hooft coupling is very small, the perturbative calculation in open string theory should be valid.}

It should be noted that we can reproduce the tachyon potential and free energy on $(N-n)$ pairs with arbitrary positive integer $n$ by changing $n$ of the diagonal components of the matrix (\ref{eq:TM}) to $\infty$. Thus, we may consider a very large matrix from the beginning, and we can treat all $N$ cases in the same framework by making such choice of the matrix.

\subsection{Free Energy near the Hagedorn Singularity}
\label{sec:Hagsing}

As mentioned in \S \ref{sec:stringgas}, the leading singularity of the free energy is at $\B = \BH$, which is called the Hagedorn singularity. This can be seen from the free energy (\ref{eq:FreeN}) by changing the variable of integration from $\tau$ to $t$, given by
\BA
  \tau = \frac{1}{t},
\label{eq:taut}
\EA
and considering the region of large $t$. Then, using the modular transformation of $\th$ functions, we obtain
\BA
  F (T, \B) &=& - \frac{16 \pi^4 N^2 \vp}{{\BH}^{p+1}}
    \int_{0}^{\infty}
      dt \ t^{\frac{p-9}{2}} \exp \left( - \frac{4 \pi \T2}{t} \right) \no \\
  && \hspace{2cm}
    \times \left[ \left(\frac{\th_3 (0 | i t)}{{\th_1}' (0 | it)} \right)^4
      \left( \th_3 \left( 0 \left| \frac{i \B^2 t}{{\BH}^2} \right. \right)
        -1 \right) \right. \no \\
  && \hspace{3cm} \left.
    - \left(\frac{\th_4 (0 | it)}{{\th_1}' (0 | it)} \right)^4
      \left( \th_4 \left( 0 \left| \frac{i \B^2 t}{{\BH}^2} \right. \right)
        -1 \right) \right].
\EA
Expanding the $\th$ functions and extracting the leading term in the large $t$ region near the Hagedorn singularity, we obtain
\BE
  F (T, \B) \simeq - \frac{4 N^2 \vp}{{\BH}^{p+1}}
    \int_{\Lambda}^{\infty} dt
      \ t^{\frac{p-9}{2}} \exp \left[ - \frac{4 \pi \T2}{t}
        - \pi \frac{\B^2 - {\BH}^2}{{\BH}^2} t \right],
\label{eq:FHag}
\EE
where we have introduced the low energy cutoff $\Lambda$. This cutoff is required, because we cannot apply the above approximation to the small $t$ region. Because we will compute the $\T2$ term of the finite temperature effective potential in the vicinity of $T=0$, we expand the free energy in $\T2$ and keep the lower order terms:
\BA
  F (T, \B) &\simeq& - \frac{4 N^2 \vp}{{\BH}^{p+1}} \int_{\Lambda}^{\infty} dt
    \ t^{\frac{p-9}{2}}
      \exp \left( - \pi \frac{\B^2 - {\BH}^2}{{\BH}^2} t \right) \no \\
  &&+ \frac{16 \pi N^2 \vp \T2}{{\BH}^{p+1}} \int_{\Lambda}^{\infty} dt
    \ t^{\frac{p-11}{2}}
      \exp \left( - \pi \frac{\B^2 - {\BH}^2}{{\BH}^2} t \right).
\label{eq:highTF}
\EA
We can rewrite (\ref{eq:highTF}) near $\B = \BH$ in terms of the incomplete $\Gamma$ function as
\BA
  F (T, \B) &\simeq&
    - \frac{2^{\a +2} {\pi}^{\a} N^2 \vp (\B - \BH)^{\a}}{{\BH}^{8- \a}}
      \ \Gamma \left( - \a \ , \ 
        2 \pi \frac{\B - \BH}{\BH} \Lambda \right) \no \\
    && + \frac{2^{\a +5} {\pi}^{\a +2} N^2 \vp (\B - \BH)^{\a +1} \T2}
      {{\BH}^{9- \a}}
        \ \Gamma \left( - \a -1 \ , \ 
          2 \pi \frac{\B - \BH}{\BH} \Lambda \right),
\label{eq:Fcut2}
\EA
where we have defined
\BE
  \a \equiv \frac{7-p}{2}.
\label{eq:adef}
\EE
We can deduce the singular part of the free energy $F_{sing}$ from (\ref{eq:Fcut2}) as follows.

\renewcommand{\descriptionlabel}[1]{\large\bfseries{#1}}
\begin{description}

\item[(a)] {\large \bf \ $p=9$  ($\a = -1$)}
\vspace{0.5cm}

When $p=9$, which means that $\a =-1$, the first argument of the incomplete $\Gamma$ function in the first term of (\ref{eq:Fcut2}) becomes $1$. Therefore we can set $\Lambda =0$. For the second term of (\ref{eq:Fcut2}), we can use the formula
\BE
  \Gamma (0,x) = -\gamma - \ln x
    - \sum_{n=1}^{\infty} \frac{(-1)^n x^n}{n \cdot n!},
\label{eq:Gammazero}
\EE
where $\gamma$ is the Euler constant. Combining these two terms, we get
\BE
  F_{sing} (T, \B) \simeq - \frac{2 N^2 \vp}{\pi {\BH}^9 (\B - \BH)}
    - \frac{16 \pi N^2 \vp \T2}{{\BH}^{10}}
      \ln \left( 2 \pi \frac{\B - \BH}{\BH} \Lambda \right).
\label{eq:FsingDd9}
\EE
To make our notation the same as that of Ref. \cite{Hotta4} we set $\Lambda = (2 \pi)^{-1}$ in this formula hereafter.

\vspace{0.5cm}
\item[(b)] {\large \bf \ $p :$ even ($\a :$ half-integer)}
\vspace{0.5cm}

When $p$ is even, the first arguments of the incomplete $\Gamma$ functions are negative half-integers. In this case, we can set $\Lambda =0$ and we obtain the singular part of the free energy as
\BA
  F_{sing} (T, \B) &\simeq&
    - \frac{4 (2 \pi)^{\a} \Gamma ( - \a) N^2 \vp}{{\BH}^{8- \a}}
      \ (\B - \BH)^{\a} \no \\
    && + \frac{8 (2 \pi)^{\a +2} \Gamma ( - \a -1) N^2 \vp \T2}{{\BH}^{9- \a}}
      \ (\B - \BH)^{\a +1}.
\label{eq:FDdeven}
\EA

\vspace{0.5cm}
\item[(c)] {\large \bf \ $p :$ odd ($\a :$ integer)}
\vspace{0.5cm}

When $p$ is odd, the first arguments of the incomplete $\Gamma$ functions are negative integers. The incomplete $\Gamma$ function whose first argument is a negative integer can be expanded as
\BE
  \Gamma (-n,x) = \frac{1}{n!} e^{-x}
    \sum_{s=1}^{n} (-1)^{s-1} (n-s)! \ x^{-n+s-1}
      + \frac{(-1)^n}{n!} \Gamma (0,x).
\EE
In our case, $\Lambda \rightarrow 0$ corresponds to $x \rightarrow 0$, and therefore the incomplete $\Gamma$ function can be approximated by the last term. Thus, the singular part of the free energy is obtained as
\BA
  F_{sing} (T, \B) &\simeq&
    - \frac{(-1)^{\a +1} 4 (2 \pi)^{\a} N^2 \vp}{\Gamma (\a +1) {\BH}^{8- \a}}
      \ (\B - \BH)^{\a}
        \ln \left( 2 \pi \frac{\B - \BH}{\BH} \Lambda \right) \no \\
    && + \frac{(-1)^{\a +2} 8 (2 \pi)^{\a +2} N^2 \vp \T2}
      {\Gamma (\a +2) {\BH}^{9- \a}}
        \ (\B - \BH)^{\a +1}
          \ln \left( 2 \pi \frac{\B - \BH}{\BH} \Lambda \right). \no \\
\label{eq:FDdodd}
\EA
We will also set $\Lambda = (2 \pi)^{-1}$ in this formula hereafter.

\end{description}

From these, we can calculate the finite temperature effective potential by using the method described in \S \ref{sec:stringgas}. For $N$ pairs of {\D{p}}, (\ref{eq:Freg}) is modified as
\BE
  - \B F_{reg} \simeq \lambda_0 N^2 \vp - \sigma_0 N^2 \vp (\B - \BH),
\label{eq:FregN}
\EE
and (\ref{eq:dos}) as
\BE
  \Omega (T,E) \simeq e^{\BH E + \lambda_0 N^2 \vp}
    \int_{C_a} \frac{d \B}{2 \pi i}
      \exp \left[ (\B - \BH) \Eb
        + F_{sing} (\B) \right],
\label{eq:dosN}
\EE
where we have defined the quantity $\Eb$ as
\BE
  \Eb \equiv E - \sigma_0 N^2 \vp.
\label{eq:EbN}
\EE
From these, we compute the finite temperature effective potential near the Hagedorn temperature.

\subsection{Non-compact Background}
\label{sec:noncompact}

In this subsection, we compute the finite temperature effective potential of multiple {\DD} pairs in a non-compact background. Although the method of calculation depends on $p$, we explicitly consider only three cases as examples, since calculation is almost the same as that in Ref. \cite{Hotta4}.

\renewcommand{\descriptionlabel}[1]{\large\bfseries{#1}}
\begin{description}

\item[(a)] {\large \bf \ $p=9$  ($\a = -1$)}
\vspace{0.5cm}

Let us first consider the case of $N$ pairs of {\D{9}}, which is the most interesting case. In this case (\ref{eq:dosN}) can be rewritten as
\BE
  \Omega (T,E) \simeq e^{\BH E + \lambda_0 N^2 \v{9}}
    \int_{C_a} \frac{d \B}{2 \pi i}
      \left( \frac{\B - \BH}{\BH} \right)^{- N^2 \Dm \v{9} \T2}
        \exp \left[ (\B - \BH) \Eb + \frac{N^2 \Cm \v{9}}{\B - \BH} \right],
\label{eq:p9dos}
\EE
where we have defined
\BA
  \Cm &=& \frac{2}{\pi {\BH}^8}, \\
  \Dm &=& - \frac{16 \pi}{{\BH}^9}.
\EA
Let us suppose that both $E$ and $E / \v{9}$ are very large. Then, the saddle point method works well, because the exponent in the integrand is very large. The result is
\BA
  \Omega (T,E) \simeq \frac{1}{2 \sqrt{\pi}}
    \left( \frac{{\BH}^2 \Eb}{N^2 \Cm \v{9}}
      \right)^{\frac{1}{2} N^2 \Dm \v{9} \T2}
        \left( \frac{N^2 \Cm \v{9}}{\Eb^3} \right)^{\frac{1}{4}} \no \\
  \times \exp \left( \BH E + \lambda_0 N^2 \v{9}
    +2 N \sqrt{\Cm \v{9} \Eb} \right).
\EA
From this $\Omega (T,E)$ and the equations (\ref{eq:Sdef}) and (\ref{eq:Bdef}), we can obtain the entropy $S(T,E)$ and the inverse temperature $\B$ as
\BA
  S(T,E) &\simeq& \frac{N^2 \Dm \v{9} \T2}{2}
    \ln \left( \frac{{\BH}^2 \Eb}{N^2 \Cm \v{9}} \right)
      - \frac{3}{4} \ln \left( \frac{\Eb}{N^{\frac{2}{3}} {\Cm}^{\frac{1}{3}}
        {\v{9}}^{\frac{1}{3}} (\delta E)^{\frac{4}{3}}} \right) \no \\
          && \hspace{4cm} + \BH E + \lambda_0 N^2 \v{9}
            + 2 N \sqrt{\Cm \v{9} \Eb},
\label{eq:p9S} \\
  \B &\simeq& \frac{N^2 \Dm \v{9} \T2}{2 \Eb}
    - \frac{3}{4 \Eb} + \BH + N \sqrt{\frac{\Cm \v{9}}{\Eb}},
\label{eq:Bmin}
\EA
respectively. It is noteworthy that the temperature is lower than the Hagedorn temperature if the energy density is very large and $T=0$. The finite temperature effective potential can be derived from (\ref{eq:Vdef}). In order to demonstrate the stability of the {\DD} system, we need only the $\T2$ term of $V(T,E)$. This term is given by
\BE
  \left[ -16 N \tau_9 \v{9}
   + \frac{8 \pi N^2 \v{9}}{{\BH}^{10}}
    \ln \left( \frac{\pi {\BH}^{10} \Eb}{2 N^2 \v{9}}
     \right) \right] \T2.
\label{eq:p9T2E}
\EE
It should be noted that the second term in the coefficient of $\T2$ is an increasing function of $E$. Because the first term is constant as long as $\v{9}$ and $\tau_9$ are fixed, the sign of the $\T2$ term changes from negative to positive at large $E$. The coefficient vanishes when
\BE
  \Eb \simeq \frac{2 N^2 \v{9}}{\pi {\BH}^{10}}
    \exp \left( \frac{2 {\BH}^{10} \tau_9}{\pi N} \right).
\label{eq:Eb9}
\EE
If we approximate (\ref{eq:Bmin}) as
\BE
  \B \simeq \BH + N \sqrt{\frac{\Cm \v{9}}{\Eb}},
\EE
we can derive the critical temperature ${\cal T}_c$ at which the coefficient vanishes. The result is
\BA
  {\cal T}_c
    &\simeq& {\BH}^{-1}
      \left[ 1+ \exp \left( - \frac{{\BH}^{10} \tau_9}
        {\pi N} \right) \right]^{-1} \no \\
  &=& {\BH}^{-1}
    \left[ 1+ \exp \left( - \frac{2^6}{g_s N} \right) \right]^{-1},
\label{eq:Tc}
\EA
where we have used the explicit formulae (\ref{eq:tension}) and (\ref{eq:BH}) in the second equality. Here, we see that this temperature is very close to the Hagedorn temperature, since $\tau_9$ is very large if the coupling of strings $g_s$ is very small. It should be noted that the critical temperature decreases as $N$ increases. Above this temperature, the coefficient of $\T2$ is positive and $T=0$ becomes the potential minimum. This implies that a phase transition occurs at the temperature ${\cal T}_c$, which is slightly below the Hagedorn temperature, and the {\D{9}} system is stable above this temperature. We can obtain the results for the single {\D{9}} pair case considered in Ref. \cite{Hotta4} by substituting $N=1$.

From (\ref{eq:EbN}) and (\ref{eq:Eb9}), we can calculate the total energy at the critical temperature as
\BA
  E_{tot} &\simeq& \sigma_0 N^2 \v{9}+ 2N \tau_9 \v{9}
    + \frac{2 N^2 \v{9}}{\pi {\BH}^{10}}
      \exp \left( \frac{2 {\BH}^{10} \tau_9}{\pi N} \right) \no \\
  &=& \sigma_0 N^2 \v{9}
    + \frac{2N \v{9}}{(2 \pi)^9 {\ap}^5 g_s}
      + \frac{2 N^2 \v{9}}{\pi (8 {\pi}^2 \ap )^5}
        \exp \left( \frac{2^7}{g_s N} \right),
\label{eq:Et9}
\EA
where we have included the brane tension energy, and we have also used the explicit formulae (\ref{eq:tension}) and (\ref{eq:BH}) in the second equality. It should be noted that for small values of $N$ and $g_s$, the total energy at the critical temperature is a decreasing function of $N$. This implies that the multiple {\D{9}} pairs are created simultaneously not in succession. Let us define $N_{min}$ as the value of $N$ that minimizes the total energy at the critical temperature. Then the $N_{min}$ pairs of {\D{9}} are first created as the energy of the system increases. If we ignore the first term of (\ref{eq:Et9}) and compute $N_{min}$ numerically, we obtain
\BE
  N_{min} \simeq \frac{45.29}{g_s}.
\label{eq:N45gN}
\EE
However, for the ideal gas approximation, we must impose the condition that the 't Hooft coupling is very small, namely,
\BE
  g_s N \ll 1,
\label{eq:gN}
\EE
and $N$ cannot be taken arbitrarily large, even if $g_s$ is very small. Equation (\ref{eq:N45gN}) does not satisfy the condition (\ref{eq:gN}). Therefore, we cannot determine the value of $N_{min}$ with a perturbative calculation. We can only say that a large number $N$ of {\D{9}} pairs are created simultaneously, because the total energy at the critical temperature is a decreasing function of $N$ as long as the 't Hooft coupling is very small.

A similar argument has been made by Danielsson, G\"{u}ijosa and Kruczenski using a calculation of the finite temperature effective potential in the case that there is only a tachyon field \cite{LowTtach}. In that case, the finite temperature effective potential is a decreasing function of $N$ if $\B$ is fixed in the canonical ensemble method, and it seems that $N$ increases without bound. However, they asserted that, given a finite amount of energy, we can determine the value of $N$ that maximizes the entropy and minimizes the finite temperature effective potential if we employ the microcanonical ensemble method, because the entropy vanishes if we use the entire energy to create {\DD} pairs or if we create no pairs at all. Our calculation demonstrates this with a calculation including the string massive mode. However, we know of no way to obtain the value of $N$ under the condition (\ref{eq:gN}).

\vspace{0.5cm}
\item[(b)] {\large \bf \ $p=8$  ($\a = - 1/2$)}
\vspace{0.5cm}

Next, let us consider the case of $N$ {\D{8}} pairs. Here, we find a result strikingly different from that in the case of the {\D{9}} system. From (\ref{eq:dosN}), the density of states is given by
\BA
  \Omega (T,E) &\simeq& e^{\BH E + \lambda_0 N^2 \v{8}}
    \int_{C_a} \frac{d \B}{2 \pi i}
      \exp \left[ (\B - \BH) \Eb
        + N^2 \Ch \v{8} (\B - \BH)^{- \frac{1}{2}} \right. \no \\
  && \hspace{6cm} \left. - N^2 \Dh \v{8} \T2 (\B - \BH)^{\frac{1}{2}}
    \right] \no \\
  &\simeq& e^{\BH E + \lambda_0 N^2 \v{8}} \int_{C_a} \frac{d \B}{2 \pi i}
    \left[ 1- N^2 \Dh \v{8} \T2 (\B - \BH)^{\frac{1}{2}} \right] \no \\
      && \hspace{3cm} \times \exp \left[ (\B - \BH) \Eb
        + N^2 \Ch \v{8} (\B - \BH)^{- \frac{1}{2}} \right],
\EA
where we have used the small $\T2$ approximation in the second equality, and we have defined
\BA
  \Ch &=& \frac{2^{\frac{3}{2}}}{{\BH}^{\frac{15}{2}}}, \\
  \Dh &=& - \frac{2^{\frac{11}{2}} {\pi}^2}{{\BH}^{\frac{17}{2}}}.
\EA
We can also use the saddle point method, which yields
\BE
  \Omega (T,E) \simeq
    \frac{N^{\frac{2}{3}} {\Ch}^{\frac{1}{3}} {\v{8}}^{\frac{1}{3}}}
      {3^{\frac{1}{2}} 2^{\frac{1}{3}} \pi^{\frac{1}{2}} \Eb^{\frac{5}{6}}}
        \exp \left( \BH E + \lambda_0 N^2 \v{8}
          + \frac{3 N^{\frac{4}{3}} {\Ch}^{\frac{2}{3}}
            {\v{8}}^{\frac{2}{3}} \Eb^{\frac{1}{3}}}
              {2^{\frac{2}{3}}}
                - \frac{N^{\frac{8}{3}} {\Ch}^{\frac{1}{3}}
                  \Dh {\v{8}}^{\frac{4}{3}} \T2}
                    {2^{\frac{1}{3}} \Eb^{\frac{1}{3}}} \right).
\EE

We can calculate the entropy $S$, the inverse temperature $\B$ and the potential $V(T,E)$ from (\ref{eq:Sdef}), (\ref{eq:Bdef}) and (\ref{eq:Vdef}), as in the {\D{9}} case. The $\T2$ term of $V(T,E)$ is given by
\BE
  \left[ -16 N \tau_8 \v{8}
    - \frac{2^{\frac{23}{3}} N^{\frac{8}{3}}
      {\v{8}}^{\frac{4}{3}}}{3 {\BH}^{12} \Eb^{\frac{1}{3}}}
        \right] \T2.
\label{eq:T2p8}
\EE
It should be noted that the second term in the coefficient of $\T2$ decreases as $E$ becomes large. Thus, the coefficient of $\T2$ remains negative for large $E$. This implies that a phase transition does not occur, unlike in the {\D{9}} case. For the $p=7,6$  ($\a = 0,1/2$) case, we can calculate the finite temperature effective potential in a similar way, again using the saddle point method.

\vspace{0.5cm}
\item[(c)] {\large \bf \ $p=3,1$  ($\a = 2,3$)}
\vspace{0.5cm}

In this case, we cannot use the saddle point method, because we cannot ignore the corrections coming from the $O( \vp (\B - \BH)^2)$ term in the expansion of the regular part of the free energy $F_{reg}$ (\ref{eq:FregN}) if we use the saddle point method, as explained in Ref. \cite{Hotta4}. We therefore adapt another type of approximation as follows.

The density of states can be obtained from (\ref{eq:dosN}) as
\BA
  \Omega (T,E) &\simeq& e^{\BH E + \lambda_0 N^2 \vp}
    \left. \int_{C_a} \frac{d \B}{2 \pi i}
      \exp \right[ (\B - \BH) \Eb \no \\
  && \left. + \left( N^2 C_{\a} (\B - \BH)^{\a}
    - N^2 D_{\a} \T2 (\B - \BH)^{\a +1} \right)
      \vp \ln \left( \frac{\B - \BH}{\BH} \right) \right], \no \\
  &&
\EA
where we have defined
\BA
  C_{\a} &=& 
    \frac{(-1)^{\a +1} 4 (2 \pi)^{\a}}{\Gamma (\a +1) {\BH}^{7- \a}}, \\
  D_{\a} &=& 
    \frac{(-1)^{\a +2} 8 (2 \pi)^{\a +2}}{\Gamma (\a +2) {\BH}^{8- \a}}.
\EA
Then, using the transformation
\BE
  z=-(\B - \BH) \Eb,
\EE
we get
\BA
  \Omega (T,E) &\simeq& - \frac{e^{\BH E + \lambda_0 N^2 \vp}}{\Eb}
    \left. \int_{C_b} \frac{dz}{2 \pi i}
      \exp \right[ -z \no \\
  && + \left. \left( (-1)^{\a} N^2 C_{\a} \frac{z^{\a}}{\Eb^{\a}}
    - (-1)^{\a +1} N^2 D_{\a} \T2 \frac{z^{\a +1}}{\Eb^{\a +1}} \right)
      \vp \left( \ln \left( \frac{z}{\BH \Eb} \right) - \pi i \right)
        \right]. \no \\
  &&
\EA
The contour $C_b$ is taken as sketched in Figure \ref{fig:com2}, where
\BE
  z_{1} =-(L' - \BH) \Eb.
\EE
We can approximate this integral with the following contour integral on both sides of the cut:
\BA
  - \int_{0}^{z_{1}} \frac{dz}{\pi} e^{-z}
    \exp \left[ \left( (-1)^{\a} N^2 C_{\a} \frac{z^{\a}}{\Eb^{\a}}
      - (-1)^{\a +1} N^2 D_{\a} \T2 \frac{z^{\a +1}}{\Eb^{\a +1}} 
        \right) \vp \ln \left( \frac{z}{\BH \Eb} \right) \right] \no \\
  \times \sin \left[ - \pi \left( (-1)^{\a} N^2 C_{\a} \frac{z^{\a}}{\Eb^{\a}}
    - (-1)^{\a +1} N^2 D_{\a} \T2 \frac{z^{\a +1}}{\Eb^{\a +1}} 
      \right) \vp \right].
\EA
Expanding the integrand here in powers of $1/ \Eb$ and taking the limit $z_{1} \rightarrow \infty$, this integral becomes
\BE
  - \frac{4 (2 \pi)^{\a}}{\pi {\BH}^{7- \a}} \frac{N^2 \vp}{\Eb^{\a}} 
    + \frac{8 (2 \pi)^{\a +2} \T2}{\pi {\BH}^{8- \a}}
      \frac{N^2 \vp}{\Eb^{\a +1}}.
\EE
Thus, the density of states is given by
\BE
  \Omega (T,E) \simeq - \frac{4 (2 \pi)^{\a} N^2 \vp}{{\BH}^{7- \a} \Eb^{\a +1}}
    e^{\BH E + \lambda_0 N^2 \vp}
      \left( 1- \frac{8 \pi^2 \T2}{\BH \Eb} \right).
\EE
Then, substituting $\Omega (T,E)$ into (\ref{eq:Sdef}), we obtain
\BA
  S(T,E) \simeq - (\a +1) \ln \left(
    \frac{\Eb {\BH}^{\frac{7- \a}{\a +1}}}
      {N^{\frac{2}{\a +1}} {\vp}^{\frac{1}{\a +1}} (\delta E)^{\frac{1}{\a +1}}}
        \right) + \BH E + \lambda_0 N^2 \vp - \frac{8 \pi^2 \T2}{\BH \Eb},
\EA
and from (\ref{eq:Bdef}), we get
\BE
  \B \simeq - \frac{\a +1}{\Eb}
    + \BH + \frac{8 \pi^2 \T2}{\BH {\Eb}^2}.
\EE
From this, we see that the temperature is higher than the Hagedorn temperature if the energy density is very large and $T=0$. This property is similar to that of a closed string gas in a non-compact background, as we see in \S \ref{sec:balance}. From the above, we can calculate the finite temperature effective potential, and its $\T2$ term is given by
\BE
  \left[ -16 N \tau_p \vp
   + \frac{16 \pi^2}{{\BH}^{2} \Eb} \right] \T2.
\EE
From this, we see that the coefficient remains negative for large $E$, and therefore no phase transition occurs. For the $p=5,4,2,0$  ($\a = 1,3/2,5/2,7/2$) cases, we can calculate the finite temperature effective potential in a similar way.
\begin{figure}
\begin{center}
$${\epsfxsize=6.5 truecm \epsfbox{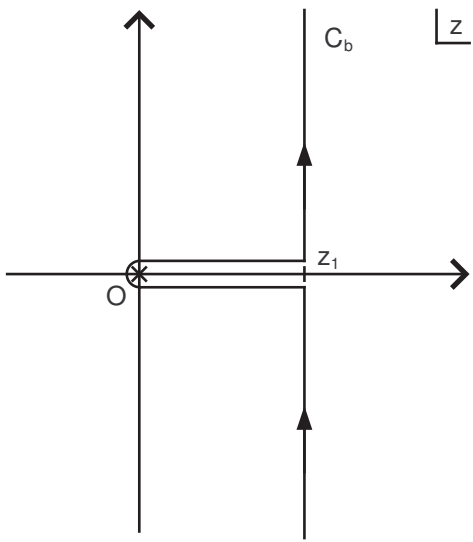}}$$
\caption{Complex $z$ plane.}
\label{fig:com2}
\end{center}
\end{figure}

\end{description}

For other cases, we only present the results, because these can be obtained with calculations similar to that given in Ref. \cite{Hotta4}. The $\T2$ term of the finite temperature effective potential is given by
\BE
  \left[ -16 N \tau_7 \v{7}
    - \frac{2^7 \pi^2 N^4 {\v{7}}^2}{{\BH}^{16} \Eb}
      \ln \left( \frac{{\BH}^8 \Eb}{4 N^2 \v{7}} \right) \right] \T2
\EE
for the {\D{7}} case,
\BE
  \left[ -16 N \tau_6 \v{6}
   + \frac{2^{17} \pi^6 N^8 {\v{6}}^4}{{\BH}^{28} \Eb^3} \right] \T2
\EE
for the {\D{6}} case,
\BE
  \left[ -16 N \tau_5 \v{5}
   + \frac{16 \pi^2}{{\BH}^{2} E'} \right] \T2
\EE
for the {\D{5}} case, where
\BE
  E' \equiv \Eb - \frac{8 \pi N^2 \v{5}}{{\BH}^6}
    \ln \left( \frac{N^2 \v{5}}{{\BH}^5} \right),
\EE
and
\BE
  \left[ -16 N \tau_p \vp
   + \frac{16 \pi^2}{{\BH}^{2} \Eb} \right] \T2
\EE
for {\D{4}}, {\D{2}} and {\D{0}} cases. From these results, we can see that the coefficients remain negative for large $E$, and therefore no phase transition occurs.  We can obtain the results for the single {\D{p}} pair case considered in Ref. \cite{Hotta4} by substituting $N=1$.

\subsection{Toroidal Background}
\label{sec:torus}

In this subsection, we generalize the calculation given in the previous subsection to the case of a toroidal background. We consider a {\D{p}} system in type II string theory compactified on a $D$-dimensional torus $T_{D}$ and assume that the {\D{p}} system is extended in the $d$-dimensional non-compact directions and in the $(p-d)$-dimensional compact directions. In this case, the free energy is given by

\newpage

\BA
  F (T, \B ,R) &=& - \frac{16 \pi^4 N^2 \v{d}}{{\BH}^{d+1}}
    \int_{0}^{\infty} \frac{d \tau}{\tau^{\frac{d+3}{2}}}
      e^{-4 \pi \T2 \tau} \prod_{I=1}^{p-d} \prod_{i=p-d+1}^{D}
        \th_3 \left( 0 \left| \frac{2 i \ap \tau}{{R_I}^2} \right. \right)
          \th_3 \left( 0 \left| \frac{2 i {R_i}^2 \tau}{\ap} \right. \right)
            \no \\
  && \hspace{2cm} \times \left[ \left(\frac{\th_3 (0 | i \tau)}
    {{\th_1}' (0 | i \tau)} \right)^4
      \left( \th_3 \left( 0 \left| \frac{i \B^2}{{\BH}^2 \tau} \right.
        \right) -1 \right) \right. \no \\
  && \hspace{4cm} - \left.
    \left( \frac{\th_2 (0 | i \tau)}{{\th_1}' (0 | i \tau)} \right)^4
      \left( \th_4 \left( 0 \left| \frac{i \B^2}{{\BH}^2 \tau} \right.
        \right) -1 \right) \right].
\label{eq:freecompactN}
\EA
This free energy is $N^2$ times that of the single pair of {\D{p}} given in (\ref{eq:freecompact}). We can obtain the free energy near the Hagedorn singularity with a calculation similar to that in the previous subsection. The result is
\BA
  F (T, \B ,R) &\simeq& - \frac{N^2 \A \vp}{\BH}
    \int_{\Lambda}^{\infty} dt
      \ t^{\frac{D+d-9}{2}} \exp \left(
        - \pi \frac{\B^2 - {\BH}^2}{{\BH}^2} t \right) \no \\
  && \times \prod_{I=1}^{p-d} \prod_{i=p-d+1}^{D}
    \th_3 \left( 0 \left| \frac{i {R_I}^2 t}{2 \ap} \right. \right)
      \th_3 \left( 0 \left| \frac{i \ap t}{2 {R_i}^2} \right. \right) \no \\
  && + \frac{4 \pi \T2 N^2 \A \vp}{\BH}
    \int_{\Lambda}^{\infty} dt
      \ t^{\frac{D+d-11}{2}} \exp \left(
        - \pi \frac{\B^2 - {\BH}^2}{{\BH}^2} t \right) \no \\
  && \times \prod_{I=1}^{p-d} \prod_{i=p-d+1}^{D}
    \th_3 \left( 0 \left| \frac{i {R_I}^2 t}{2 \ap} \right. \right)
      \th_3 \left( 0 \left| \frac{i \ap t}{2 {R_i}^2} \right. \right),
\label{eq:FHag2}
\EA
where we have defined
\BE
  \A = \frac{{\ap}^{d-p+ \frac{D}{2}}}
    {2^{\frac{D}{2} -2} {\BH}^{d} \prod_{i=p-d+1}^{D} R_i}
\EE
and
\BE
  \vp = \v{d} \prod_{I=1}^{p-d} R_I.
\EE
This free energy is invariant under the T-duality transformation given in (\ref{eq:Tdualmn}) and (\ref{eq:Tdualnm}). We only need to investigate the region in which $R \geq \sqrt{\ap}$.

Let us first consider the case that all the radii are close to the string scale, $\sqrt{\ap}$. In this case, the Hagedorn singularity is dominant, and we need to consider only the terms
\BA
  F (T, \B ,R) &\simeq& - \frac{N^2 \A \vp}{\BH}
    \int_{\Lambda}^{\infty} dt
      \ t^{\frac{D+d-9}{2}} \exp \left(
        - \pi \frac{\B^2 - {\BH}^2}{{\BH}^2} t \right) \no \\
  && + \frac{4 \pi \T2 N^2 \A \vp}{\BH}
    \int_{\Lambda}^{\infty} dt
      \ t^{\frac{D+d-11}{2}} \exp \left(
        - \pi \frac{\B^2 - {\BH}^2}{{\BH}^2} t \right).
\label{eq:FHagRsmall}
\EA
This free energy is proportional to that in the non-compact background case, (\ref{eq:highTF}), if we replace $(D+d)$ by $p$. From this, we can conjecture that a phase transition occurs only when $D+d=9$. In the $D+d=9$ case, the $\T2$ term of the finite temperature effective potential is given by
\BE
  \left[ -16 N \tau_p \vp
   + \frac{2 \pi N^2 \A \vp}{\BH}
    \ln \left( \frac{2 \pi \BH \Eb}{N^2 \A \vp}
     \right) \right] \T2.
\EE
It should be noted that the second term in the coefficient of $\T2$ is an increasing function of $E$. Because the first term is constant as long as $\vp$ and $\tau_p$ are fixed, the sign of the $\T2$ term changes from negative to positive at large $E$. This means that a phase transition occurs near the Hagedorn temperature. The coefficient vanishes when
\BE
  \Eb \simeq \frac{N^2 \A \vp}{2 \pi \BH}
    \exp \left( \frac{8 \BH \tau_p}{\pi N \A} \right).
\label{eq:EDd9}
\EE
The critical temperature ${\cal T}_c$ and the total energy $E_{tot}$ at the critical temperature are given by
\BE
  {\cal T}_c
    \simeq {\BH}^{-1}
      \left[ 1+ \exp \left( - \frac{4 \BH \tau_p}
        {\pi N \A} \right) \right]^{-1}
\label{eq:TcR}
\EE
and
\BE
  E_{tot} \simeq \sigma_0 N^2 \v{p} + 2N \tau_p \v{p}
    + \frac{N^2 \A \vp}{2 \pi \BH}
      \exp \left( \frac{8 \BH \tau_p}{\pi N \A} \right),
\label{eq:Et9R}
\EE
respectively. The total energy at the critical temperature is also a decreasing function of $N$ as long as the 't Hooft coupling is very small, as in the case of {\D{9}} in a non-compact background. This implies that a large number $N$ of {\D{p}} pairs are created simultaneously.

In the $D+d \leq 8$ case, the $\T2$ term in the finite temperature effective potential is given by
\BE
  \left[ -16 N \tau_p \vp
    - \frac{2^5 \pi^2 N^{\frac{8}{3}} {\A}^{\frac{4}{3}} {\vp}^{\frac{4}{3}}}
      {3 {\BH}^{\frac{4}{3}} \Eb^{\frac{1}{3}}}
     \right] \T2
\EE
for $D+d = 8$,
\BE
  \left[ -16 N \tau_p \vp
   - \frac{8 \pi^2 N^4 {\A}^2 {\vp}^2}{\BH \Eb}
     \ln \left( \frac{\BH \Eb}{N^2 \A \vp} \right) \right] \T2
\EE
for $D+d = 7$,
\BE
  \left[ -16 N \tau_p \vp
   + \frac{2^{10} \pi^6 N^8 {\A}^4 {\vp}^4}{6 {\BH}^4 \Eb^3} \right] \T2
\EE
for $D+d = 6$,
\BE
  \left[ -16 \tau_p \vp
   + \frac{16 \pi^2}{{\BH}^{2} E'} \right] \T2
\EE
for $D+d = 5$, where
\BE
  E' \equiv \Eb - \frac{2 \pi N^2 \A \vp}{\BH} \ln \left( N^2 \A \vp \right),
\EE
and
\BE
  \left[ -16 N \tau_p \vp
   + \frac{16 \pi^2}{{\BH}^{2} \Eb} \right] \T2
\EE
for $D+d \leq 4$. From these results we can see that the coefficients remain negative for large $E$, and therefore no phase transition occurs in these cases. We can obtain the results for the single {\D{p}} pair case considered in Ref. \cite{Hotta5} by substituting $N=1$.

\begin{figure}
\begin{center}
$${\epsfxsize=6.5 truecm \epsfbox{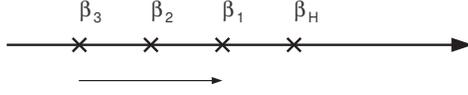}}$$
\caption{Radius dependent singularities.}
\label{fig:com3}
\end{center}
\end{figure}

Next, let us consider the case that the radii $R_{\db}$ in the $\db$-dimensional directions transverse to the {\D{p}} system are much larger than the string scale, $\sqrt{\ap}$. In this case, the radius dependent singularities exist near the Hagedorn singularity, as sketched in Figure \ref{fig:com3}, and we must consider the radius dependent terms. Because one of the infinite product of radius dependent $\th$ functions in (\ref{eq:FHag2}) can be rewritten as
\BE
  \prod_{i=p-d+1}^{p-d+ \db}
    \th_3 \left( 0 \left| \frac{i \ap t}{2 {R_{\db}}^2} \right. \right)
      = \sum_{n_i =- \infty}^{\infty} \exp \left( - \sum_{i=p-d+1}^{p-d+ \db}
        \frac{\pi {n_i}^2 \ap}{2 {R_{\db}}^2} t \right),
\EE
the free energy can be approximated as
\BA
  F (T, \B ,R) &\simeq& - \frac{N^2 \A \vp}{\BH}
    \int_{\Lambda}^{\infty} dt
      \sum_{n_i =- \infty}^{\infty} \exp \left(
        - \pi \frac{\B^2 - {\Bn}^2}{{\BH}^2} t \right) \no \\
  && + \frac{4 \pi \T2 N^2 \A \vp}{\BH}
    \int_{\Lambda}^{\infty} \frac{dt}{t}
      \sum_{n_i =- \infty}^{\infty} \exp \left(
        - \pi \frac{\B^2 - {\Bn}^2}{{\BH}^2} t \right),
\label{eq:FHagRlarge}
\EA
where we have defined
\BE
  {\Bn}^2 \equiv {\BH}^2 \left( 1- \sum_{i=p-d+1}^{p-d+ \db}
    \frac{{n_i}^2 \ap}{2 {R_{\db}}^2} \right).
\label{eq:Bn}
\EE
From this we see that singularities exist at $\B = \Bn$ on the real axis of the complex $\B$-plane, and they approach the Hagedorn singularity $\B = \BH$ as $R_{\db}$ increases.

Even if $R_{\db}$ is much larger than the string scale, a phase transition occurs only in the $D+d=9$ case. The coefficient of the $\T2$ term of the finite temperature effective potential becomes positive only when the density of states is dominated by the contribution from the Hagedorn singularity and we can use the saddle point method to compute the density of states, as in the $N=1$ case, which we considered in Ref. \cite{Hotta5}. These conditions are realized if $\Eb$ satisfies the relation
\BE
  \Eb \gg \max \left( \sigma_0 N^2 \vp,
    \frac{{\ap}^{\frac{p- \db-1}{2}} {R_{\db}}^{\db}}{N^2 \vp} ,
      \frac{N^2 {R_{\db}}^{4- \db} \vp}{{\ap}^{\frac{5+p- \db}{2}}} \right).
\label{eq:regionA}
\EE
Then the finite temperature effective potential has the same form as that in the case that $\db =0$ and $D+d=9$. A phase transition occurs when $\Eb$ is given by (\ref{eq:EDd9}), which satisfies (\ref{eq:regionA}) as long as the 't Hooft coupling is very small. The critical temperature is given by (\ref{eq:TcR}). Because $A$ depends on $R_{\db}$, the total energy at the critical temperature is an increasing function of $R_{\db}$ as long as the 't Hooft coupling is very small.

\section{Non-BPS D-brane}
\label{sec:nonBPS}

We briefly study the non-BPS D-brane case in this section. If we consider $N$ non-BPS D-branes, the spectrum of open strings contains a tachyon field in the adjoint representation of the $U(N)$ gauge group \cite{nonBPSD}. The tension of a single non-BPS Dp-brane is given by $\sqrt{2} \tau_p$ \cite{nonBPSten}, and the tree level tachyon potential of $N$ non-BPS D-branes is given by \cite{tachyon2} \cite{TakaTeraUe}
\BE
  V(T) = \sqrt{2} N \tau_p \vp \exp (-2 T^2),
\EE
where $T$ is a real variable, and we have chosen the matrix as (\ref{eq:TM}). This is $1/ \sqrt{2}$ of the tachyon potential in the {\D{p}}, case with $\T2$ replaced by $T^2 /4$. The one-loop free energy is given by half of that in the {\D{p}}, case with $\T2$ replaced by $T^2 /4$ \cite{1loopAO}. The free energy of open strings on $N$ non-BPS D-branes, for example, in a non-compact background is given by
\BA
  F (T, \B) &=& - \frac{8 \pi^4 N^2 \vp}{{\BH}^{p+1}}
    \int_{0}^{\infty} \frac{d \tau}{{\tau}^{\frac{p+3}{2}}}                          \exp \left( - \pi T^2 \tau \right) \no \\
  && \hspace{2cm} \times \left[ \left(\frac{\th_3 (0 | i \tau)}
    {{\th_1}' (0 | i \tau)} \right)^4
      \left( \th_3 \left( 0 \left| \frac{i \B^2}{{\BH}^2 \tau} \right.
        \right) -1 \right) \right. \no \\
  && \hspace{4cm} - \left.
    \left( \frac{\th_2 (0 | i \tau)}{{\th_1}' (0 | i \tau)} \right)^4
      \left( \th_4 \left( 0 \left| \frac{i \B^2}{{\BH}^2 \tau} \right.
        \right) -1 \right) \right].
\EA
This free energy can be obtained from the proper time form of the free energy (\ref{eq:propertime}) by multiplying by $N^2$ and substituting the mass spectra (\ref{eq:massNS}) and (\ref{eq:massR}) with $2 \T2$ replaced by $T^2 /2$. The factor $N^2$ can be interpreted in the context of open string theory as follows. There are $N^2$ types of strings projected by the ordinary GSO projection and $N^2$ types of strings projected by the opposite GSO projection \cite{nonBPSD}. Therefore, instead of employing the GSO projection, we may multiply $N^2$ as an overall factor. From these considerations, we see that we can obtain the results in the case of non-BPS D-branes from those in the case of {\DD} pairs by replacing $N$ by $N/ \sqrt{2}$. Below, we give the results only for the cases of non-BPS D9-branes in a non-compact background and of non-BPS Dp-branes with $D+d=9$ in a toroidal background, in which a phase transition occurs.

In the case of $N$ non-BPS D9-branes in a non-compact background, the $T^2$ term of the finite temperature effective potential is given by
\BE
  \frac{1}{4} \left[ -8 \sqrt{2} N \tau_9 \v{9}
   + \frac{4 \pi N^2 \v{9}}{{\BH}^{10}}
    \ln \left( \frac{\pi {\BH}^{10} \Eb}{N^2 \v{9}}
     \right) \right] T^2.
\EE
The critical temperature ${\cal T}_c$ and the total energy $E_{tot}$ at the critical temperature are given by\footnote{We have not divided the first term on the right hand of (\ref{eq:EtotnonBPS}) by $\sqrt{2}$, because we have defined $\sigma_0$ by (\ref{eq:FregN}) and $\Eb$ by (\ref{eq:EbN}).}
\BE
  {\cal T}_c
    \simeq {\BH}^{-1}
      \left[ 1+ \exp \left( - \frac{\sqrt{2} {\BH}^{10} \tau_9}
        {\pi N} \right) \right]^{-1},
\EE
and
\BE
  E_{tot} \simeq \sigma_0 N^2 \v{9}+ \sqrt{2} N \tau_9 \v{9}
    + \frac{N^2 \v{9}}{\pi {\BH}^{10}}
      \exp \left( \frac{2 \sqrt{2} {\BH}^{10} \tau_9}{\pi N} \right).
\label{eq:EtotnonBPS}
\EE
In the case of $N$ non-BPS Dp-branes with $D+d=9$ in the toroidal background, the $T^2$ term of the finite temperature effective potential is given by
\BE
  \frac{1}{4} \left[ -8 \sqrt{2} N \tau_p \vp
   + \frac{\pi N^2 \A \vp}{\BH}
    \ln \left( \frac{4 \pi \BH \Eb}{N^2 \A \vp}
     \right) \right] T^2.
\EE
The critical temperature ${\cal T}_c$ and the total energy $E_{tot}$ at the critical temperature are given by
\BE
  {\cal T}_c
    \simeq {\BH}^{-1}
      \left[ 1+ \exp \left( - \frac{4 \sqrt{2} \BH \tau_p}
        {\pi N \A} \right) \right]^{-1},
\EE
and
\BE
  E_{tot} \simeq \sigma_0 N^2 \v{p} + \sqrt{2} N \tau_p \v{p}
    + \frac{N^2 \A \vp}{4 \pi \BH}
      \exp \left( \frac{8 \sqrt{2} \BH \tau_p}{\pi N \A} \right).
\EE
In these two cases, the total energy at the critical temperature is also a decreasing function of $N$ as long as the 't Hooft coupling is very small, as in the {\D{p}} case. This implies that a large number $N$ of non-BPS D-branes are created simultaneously by a phase transition in these cases. No phase transition occurs in the other cases.

\section{Thermodynamic Balance}
\label{sec:balance}

To this point, we have ignored closed strings. We need to consider not only open strings but also closed strings if we wish to deal with the entire the system of strings in type II string theory. Abel, Barbon, Kogan and Rabinovici have investigated the finite temperature system of BPS D-branes on a 9-dimensional torus and treated the thermodynamic balance between open strings on the BPS D-branes and closed strings in the bulk \cite{Thermo}. In a similar way, in this section we treat the thermodynamic balance between open strings and closed strings in the case that {\DD} pairs or non-BPS D-branes become stable near the Hagedorn temperature.

We begin by reviewing the thermodynamics of closed strings in a $10$-dimensional non-compact spacetime. We can reach the Hagedorn temperature for closed strings by supplying a finite amount of energy if we consider only closed strings \cite{limiting}. Based on this fact, it has been said that the Hagedorn temperature in the closed string case is associated with a phase transition, in analogy to the deconfining transition in QCD. Sathiapalan \cite{Sa}, Kogan \cite{Ko} and Atick and Witten \cite{AW} have argued that this phase transition occurs because the `winding modes' of the Euclidean time direction become tachyonic above the Hagedorn temperature, and these tachyon fields condense towards the potential minimum. This phase transition is called the `Hagedorn transition'. We do not yet know where the minimum of this tachyon potential is and in what kind of backgrounds the system becomes stable. Thus, there is a possibility that we cannot apply the ideal string gas approximation to closed strings above the Hagedorn energy density. However, we temporarily assume that we can treat closed strings as an ideal gas, because it is expected that closed strings cannot reach the Hagedorn energy density if we consider thermodynamic balance between closed strings and open strings, as we now explain.

We can investigate the thermodynamic balance of open strings and closed strings by considering the fact that the entropy becomes maximal in the equilibrium state. Let us fix the total energy $E_{tot}$ and consider the variation of the total entropy $S_{tot}$. $E_{tot}$ is given by
\BE
  E_{tot} = E_o + E_c + E_b,
\EE
where we have denoted the energy of open strings by $E_o$, that of closed strings by $E_c$, and that of the brane tension by $E_b$. $S_{tot}$ is given by
\BE
  S_{tot} = S_o + S_c,
\EE
where we have denoted the entropy of open strings by $S_o$ and that of closed strings by $S_c$. By using the definition of the temperature (\ref{eq:Bdef}) in the microcanonical ensemble framework, we can calculate the partial derivative of $S_{tot}$ with respect to $E_c$ as
\BE
  \frac{\p S_{tot}}{\p E_c}
    = \frac{\p S_c}{\p E_c} - \frac{\p S_o}{\p E_o}
      = \frac{1}{{\cal T}_{closed}} - \frac{1}{{\cal T}_{open}},
\label{eq:pSpE}
\EE
where we have used the fact that $E_b$ is a constant and the approximation that $E_{tot}$ is fixed in the first equality. When $S_{tot}$ realizes its maximum and the system becomes the equilibrium state, this partial derivative vanishes. Therefore, the thermodynamic balance condition is given by
\BE
  {\cal T}_{open} = {\cal T}_{closed}.
\label{eq:TTequiv}
\EE
That is, thermodynamic balance is realized when the temperatures of the two types of strings become the same. If the temperature of one component is higher than that of other, then energy flows from the hotter to the cooler.

Let us consider, for example, the {\D{9}} system in a non-compact background under the condition that the energy is sufficiently large to create {\D{9}} pairs. The entropy of open strings is given by (\ref{eq:p9S}), and it can be approximated for large $\Eb$ as
\BE
  S_o \simeq \BH E_o + 2 N \sqrt{\Cm \v{9} \Eb_o}
\EE
at $T=0$, where $\Eb_o = E_o - \sigma_0 N^2 \vp$. The temperature is given by
\BE
  {\cal T}_{open} \simeq
    \left[ \BH + N \sqrt{\frac{\Cm \v{9}}{\Eb_o}} \ \right]^{-1},
\EE
which is lower than the Hagedorn temperature, i.e., ${\cal T}_{open} < {\cal T}_{H}$. We need an infinite energy to reach the Hagedorn temperature, and this is a `limiting temperature' \cite{limiting} in the case of an open string on the {\D{9}} pair. On the other hand, if we calculate the entropy of closed strings that have an energy density that is much larger than the Hagedorn energy density in the microcanonical ensemble method, we obtain \cite{Tan2}\footnote{Because the free energy of a closed string gas behaves as if it were the case that $\a =5$, we can calculate the entropy and the temperature as in case (c) of \S \ref{sec:noncompact}.}
\BE
  S_c \simeq \BH E_c - \frac{11}{2}
    \ln \left( \frac{{\ap}^{\frac{27}{22}} {\Eb_c}'}
      {{\v{9}}^{\frac{2}{11}} \delta {E_c}^{\frac{2}{11}}} \right),
\EE
where ${\Eb_c}' = E_c - \sigma_0 \v{9}$ and $\delta E_c$ represents the energy fluctuation. The temperature is given by
\BE
  {\cal T}_{closed} \simeq
    \left[ \BH - \frac{11}{2} \frac{1}{{\Eb_c}'} \right]^{-1},
\EE
which is higher than the Hagedorn temperature, i.e., ${\cal T}_{closed} > {\cal T}_{H}$, and the temperature approaches the Hagedorn temperature from above. Therefore, the thermodynamic balance condition (\ref{eq:TTequiv}) is not satisfied as long as the closed strings have an energy density larger than the Hagedorn energy density, and the energy flows from the closed strings to the open strings. To put it another way, by substituting these temperatures into (\ref{eq:pSpE}), we can conclude that $S_{tot}$ is a monotonically decreasing function of $E_c$ throughout the entire Hagedorn regime. These facts imply that energy flows from closed strings to open strings before the closed strings realize the Hagedorn energy density. The closed strings cannot realize the Hagedorn energy density, and its temperature does not reach the Hagedorn temperature. Therefore, if we consider thermodynamic balance, open strings dominate the total energy of strings as long as this total energy is large enough to create {\D{9}} pairs.

If we consider the thermodynamic balance of open strings on the {\D{p}} pairs with $D+d=9$ and closed strings in a toroidal background, we can also show that open strings dominate the total energy of strings \cite{Thermo}, although there are cases in which the temperature of the closed strings is lower than the Hagedorn temperature even if they have an energy density larger than the Hagedorn energy density. This is because open strings in these cases have a larger entropy than closed strings if they have the same energy, and energy flows from closed strings to open strings in order to maximize the total entropy \cite{Thermo}.\footnote{D9-branes and non-BPS D9-branes in a non-compact background and {\D{p}} pairs and non-BPS Dp-branes with $D+d=9$ in a toroidal background are classified into ${\bf L} [-1]$ in Ref. \cite{Thermo}. ${\bf L} [-1]$ systems dominate the string energy, because they have the largest entropy under the condition that each type of string have the same energy. They are the most `favorable'.} We can derive a similar conclusion in the non-BPS D-brane case. Therefore, for all cases in which a phase transition occurs, open strings dominate the total energy, and closed strings have only a small energy. These facts indicate that the open string degrees of freedom are very important near the Hagedorn temperature.

\section{Annulus Boundary Action}
\label{sec:annulus}

In this section, we consider an annulus-type world sheet as an example of a different choice of the Weyl factors. In the single {\D{p}} pair case, the natural choice of the Weyl factors of this world sheet gives the following form of the boundary action \cite{1loopann}:\footnote{We can choose another type of Weyl factor for the annulus world sheet \cite{1loopsym} In this case, we obtain the results similar to those for the choice (\ref{eq:Sann}).}
\BE
  S_{ann} = \frac{1}{2 \pi} \int_{0}^{2 \pi} d \theta \ (1+q) \T2,
\label{eq:Sann}
\EE
where we have set the radius the inner boundary to $q$ and that of the outer boundary to $1$.\footnote{If we choose the radii as $qr$ and $r$, with a positive real number $r$, respectively, we can absorb $r$ through the redefinition of $T$.} We call this action the `annulus boundary action'. The one-loop partition function is given by \cite{1loopAO} \cite{1loopann}
\BA
  Z_{1} &=& \frac{{\pi}^{\frac{7-p}{2}} \ i \vp}{{\BH}^{p+1}}
    \int_{0}^{1} \frac{dq}{q^2} \exp \left[ -(1+q) \T2 \right]
      \left( - \ln q \right)^{\frac{p-9}{2}}
        \prod_{n=1}^{\infty} (1-q^{2n} )^{-8} \no \\
  && \hspace{3cm} \times \left[ \prod_{n=1}^{\infty} (1+q^{2n-1} )^8
    - \prod_{n=1}^{\infty} (1-q^{2n-1} )^8 \right].
\EA
If we change the variable of integration from $q$ to $\tau$, given by
\BE
  q = \exp \left( - \frac{\pi}{\tau} \right),
\EE
we can obtain a form of the one-loop amplitude that is similar to that in the case of the cylinder boundary action (\ref{eq:AOoneloopAmp}). Explicitly, we have
\BA
  Z_{1} = \frac{16 \pi^4 i \vp}{{\BH}^{p+1}}
    \int_{0}^{\infty} \frac{d \tau}{\tau^{\frac{p+3}{2}}}
      \exp \left[ -(1+e^{- \frac{\pi}{\tau}}) \T2 \right]
        \left[ \left( \frac{\th_3 (0 | i \tau)}{{\th_1}' (0 | i \tau)} \right)^4
          - \left(\frac{\th_2 (0 | i \tau)}{{\th_1}' (0 | i \tau)}
            \right)^4 \right].
\EA
If we compute the one-loop free energy using the Matsubara method, we obtain
\BA
  F (T, \B) &=& - \frac{16 \pi^4 \vp}{{\BH}^{p+1}}
    \int_{0}^{\infty} \frac{d \tau}{\tau^{\frac{p+3}{2}}}
      \exp \left[ -(1+e^{- \frac{\pi}{\tau}}) \T2 \right] \no \\
  && \hspace{2cm} \times \left[ \left(\frac{\th_3 (0 | i \tau)}
    {{\th_1}' (0 | i \tau)} \right)^4
      \left( \th_3 \left( 0 \left| \frac{i \B^2}{{\BH}^2 \tau} \right.
        \right) -1 \right) \right. \no \\
  && \hspace{4cm} - \left.
    \left( \frac{\th_2 (0 | i \tau)}{{\th_1}' (0 | i \tau)} \right)^4
      \left( \th_4 \left( 0 \left| \frac{i \B^2}{{\BH}^2 \tau} \right.
        \right) -1 \right) \right].
\EA
We cannot obtain this free energy by substituting any form of the mass spectra into the proper time form of the free energy (\ref{eq:propertime}). Therefore we cannot explain this free energy in terms of a simple mass shift, like (\ref{eq:massNS}) and (\ref{eq:massR}). In order to see the singular behavior of the free energy near the Hagedorn singularity, let us change the variable of integration from $\tau$ to $t$, according to (\ref{eq:taut}), and extract the leading term in large $t$ region near the Hagedorn singularity. We then obtain
\BA
  F (T, \B) &\simeq& - \frac{4 \vp}{{\BH}^{p+1}} e^{- \T2}
    \int_{\Lambda}^{\infty} dt \ t^{\frac{p-9}{2}}
      \exp \left( - \pi \frac{\B^2 - {\BH}^2}{{\BH}^2} t \right) \no \\
  &\simeq& - \frac{2^{\a +2} {\pi}^{\a} \vp e^{- \T2}
    (\B - \BH)^{\a}}{{\BH}^{8- \a}}
      \ \Gamma \left( - \a \ , \ 
        2 \pi \frac{\B - \BH}{\BH} \Lambda \right),
\EA
where $\Lambda$ is the low energy cutoff and $\a$ is defined by (\ref{eq:adef}). It should be noted that the contribution of the tachyon $T$ comes from only one boundary. This is very unnatural and a very different situation from that for the cylinder boundary action, which gives the same contribution to both boundaries. If we expand this free energy in $\T2$, the power of $(\B - \BH)$ in the $\T2$ term is different from that in (\ref{eq:Fcut2}). Therefore, we obtain a different result than in the cylinder boundary action case. The change from the power in the cylinder boundary action case to that in the annulus boundary action case effectively corresponds to the replacement of $\a$ by $(\a -1)$, so that a phase transition occurs in the {\D{p}} case with $p \leq 7$. We only present the results for $N$ {\D{p}} pairs in these cases. In the {\D{9}} case, the $\T2$ term of the finite temperature effective potential is given by
\BE
  \left[ -16 N \tau_9 \v{9}
   + \frac{1}{{\BH}^5}
     \sqrt{\frac{N^2 \v{9} \Eb}{2 \pi}} \right] \T2,
\EE
and the total energy at the critical temperature is
\BE
  E_{tot} \simeq \sigma_0 N^2 \v{9} + 2N \tau_9 \v{9}
    + 2^9 \pi {\BH}^{10} {\tau_9}^2 \v{9}.
\EE
In the {\D{8}} case, the $\T2$ term is given by
\BE
  \left[ -16 N \tau_8 \v{8}
   + \frac{2^{\frac{7}{3}} N^{\frac{4}{3}} {\v{8}}^{\frac{2}{3}}
     {\Eb}^{\frac{1}{3}}}{3 {\BH}^6} \right] \T2,
\EE
and the total energy at the critical temperature is
\BE
  E_{tot} \simeq \sigma_0 N^2 \v{8} + 2N \tau_8 \v{8}
    + \frac{2^5 3^3 {\BH}^{18} {\tau_8}^3 \v{8}}{N}.
\EE
In the {\D{7}} case, the $\T2$ term is given by
\BE
  \left[ -16 N \tau_7 \v{7}
   + \frac{4 N^2 \v{7}}{{\BH}^8}
     \ln \left( \frac{{\BH}^8 \Eb}{4 N^2 \v{7}} \right)
       \right] \T2,
\EE
and the total energy at the critical temperature is
\BE
  E_{tot} \simeq \sigma_0 N^2 \v{7} + 2N \tau_7 \v{7}
    + \frac{4 N^2 \v{7}}{{\BH}^8} \exp \left( \frac{4 {\BH}^8 \tau_7}{N} \right).
\EE
The total energy at the critical temperature is a increasing function of $N$ in the {\D{9}} case, while it is a decreasing function of $N$ in the {\D{8}} and {\D{7}} cases, as long as the 't Hooft coupling is very small, as in the cylinder boundary action case. However, we cannot obtain the value of $N_{min}$ for {\D{8}} and {\D{7}} pairs in the weak coupling region (\ref{eq:gN}). Thus we cannot determine which {\D{p}} pairs are created first as the total energy increases. In the case of lower-dimensional {\D{p}} pairs, no phase transition occurs. In a toroidal background, a phase transition occurs only in the $D+d \geq 7$ case. We obtain a similar results in the non-BPS D-brane case.

The property common to the cylinder boundary action case and the annulus boundary action case is that only the higher-dimensional branes are created by the phase transition in a non-compact background. Whether a phase transition occurs or not depends on the power of $(\B - \BH)$ of the $\T2$ term in the free energy near the Hagedorn singularity. It seems that this power in a non-compact background is a decreasing function of $p$ for any choice of the Weyl factors. In the cylinder boundary action case, for example, this power is $(\a -1)$, as we can see from (\ref{eq:Fcut2}), and $\a$ is defined as (\ref{eq:adef}), so that the power is a decreasing function of $p$. This property originates from the contribution from the momentum modes of open strings, which can be taken in the $p$-dimensional spatial directions on the {\D{p}} pairs. Therefore, it is expected that only higher-dimensional branes are created by the phase transition for any choice of the Weyl factors. It is also expected that only branes with large $(D+d)$ are created by the phase transition in a toroidal background for any choice of the Weyl factors.

\section{Conclusion and Discussion}
\label{sec:conclusion}

In this paper, we have discussed the thermodynamic properties of {\D{p}} pairs and non-BPS D-branes in a constant tachyon background. We generalized the argument given in Refs. \cite{Hotta4} and \cite{Hotta5} to that for $N$ {\D{p}} pairs and evaluated the $\T2$ term of the finite temperature effective potential by using the microcanonical ensemble method near the Hagedorn temperature. In the {\D{9}} case in a non-compact background and in the $D+d=9$ case in a toroidal background, a phase transition occurs slightly below the Hagedorn temperature. The total energy at the critical temperature is a decreasing function of $N$ as long as the 't Hooft coupling is very small in these cases. This implies that a large number $N$ of {\DD} pairs are created simultaneously. However, we cannot determine the value of $N_{min}$ with a perturbative calculation. In order to determine it, we must perform a non-perturbative calculation based on, for example, the matrix model \cite{matrix} or the IIB matrix model \cite{IIBmatrix}. The K-matrix model may also be useful, as it explicitly contains the tachyon field \cite{Kmatrix}. Such a phase transition does not occur in the other cases. We have also investigated the behavior of a finite temperature effective potential on non-BPS D-branes and obtained similar results.

The following is another reason that we need to perform a non-perturbative calculation. We have treated the phase transition near the Hagedorn temperature by calculating the $|T|^2$ term of the finite temperature effective potential up to the one-loop effect of open strings. By analogy with the Hagedorn transition in closed string theory, it seems that we cannot ignore the higher-loop contribution to the finite temperature effective potential if the total energy is larger than the brane tension energy. However, it is not clear whether we can ignore the higher-loop effect or not, because we are considering the $|T|^2$ term of the finite temperature effective potential of open strings. Actually, we can ignore the one-loop contribution in comparison with the tree level contribution in the lower-dimensional brane cases, as we have seen, for example, in \S \ref{sec:noncompact}. Thus, we must compute the higher-loop effect explicitly in order to verify our calculations up to one-loop. However, we have not succeeded in computing the higher-loop contribution by using BSFT to this time, because we do not know the natural choice of the Weyl factors in the higher-loop case, unlike in the one-loop case. We leave the non-perturbative calculation for the future.

We have investigated the thermodynamic balance between the open strings on these branes and the closed strings in the bulk in the ideal gas approximation, and we have found that the open strings dominate the total energy. Thus, we conclude that the open string degrees of freedom are very important near the Hagedorn temperature. Strictly speaking, however, we must consider the Boltzmann equation for open strings and closed strings in order to elucidate the thermodynamic balance, even if we can treat the system perturbatively \cite{HotSoup}.

The calculations given here are based on the cylinder boundary action, which was proposed by Andreev and Oft \cite{1loopAO}. As mentioned in \S \ref{sec:stringgas}, there is a problem in choosing the Weyl factors in the two boundaries of the one-loop world sheet. The cylinder boundary action is natural because with it, both sides of the world sheet are treated on an equal footing, and its low energy part coincides with that of the tachyon field model \cite{tachyon2} \cite{TakaTeraUe} \cite{tachyon1}. If we calculate the finite temperature effective potential in the annulus boundary action case, we obtain results that differ from those in the cylinder boundary action case. However, in a non-compact background, a phase transition also occurs only in the case of higher-dimensional branes. As mentioned in \S \ref{sec:annulus}, it is expected that only higher-dimensional branes in a non-compact background and branes with large $(D+d)$ in a toroidal background are created by the phase transition, with any choice of the Weyl factors.

It is possible that the {\DD} pairs and the non-BPS Dp-branes are stable in the early universe, as the temperature is extremely high in this period. Then the universe expands in an inflationary manner, because the tension energy of these branes can provide an effective cosmological constant. A considerable number of studies have been made on brane inflation \cite{inflation1} \cite{inflation2} \cite{inflation3} (for a review see, e.g., Ref. \cite{Gibbons}). We must consider gravity coupled with the tachyon field, including the finite temperature tachyon potential, in order to analyze a time dependent background. If we use the finite temperature effective potential we obtained, it is natural to treat this background in the framework of BSFT. The time dependent background at zero temperature was calculated by Sugimoto and Terashima in the framework of BSFT \cite{BSFTcosmo}. It would be interesting to generalize their calculation to the finite temperature case. It is expected that, even if these branes are stable initially, they become unstable, and the tachyon starts to roll down from $T=0$ \cite{roll}, because the energy density decreases as the universe expands.

It is noteworthy that spacetime-filling branes, such as the {\D{9}} pairs and non-BPS D9-branes, are created at sufficiently high energy not only in a non-compact background but also in a toroidal background, since these branes always satisfy $D+d=9$. These spacetime-filling branes are very advantageous in the sense that all the lower-dimensional D-branes are realized as topological defects through tachyon condensation from the spacetime-filling branes \cite{Ktheory1} \cite{Ktheory2}. It would be interesting to consider the possibility that our `brane world' forms as a topological defect in the process of tachyon condensation in a cosmological context \cite{inflation3}.

Finally, the phase transition to spacetime-filling branes is reminiscent of the phase transition in the Plank solid model of Schwarzschild black holes \cite{Hotta3}. It might be interesting to study black holes as multiple spacetime-filling branes.

\section*{Acknowledgements}

The author would like to thank S. Sugimoto for discussions and comments on the manuscript. He also would like to thank K. Ito, K. Hashimoto, S. Nakamura, D. Tomino, K. Nagami, K. Takahashi and colleagues at Kyoto University and Tokyo Institute of Technology for useful discussions. He appreciates the members of the Yukawa Institute for Theoretical Physics at Kyoto University. Discussions during the YITP workshop YITP-W-03-07, ``Quantum Field Theory 2003,'' were useful in completing this work.

\end{document}